\documentclass[aps, prd, preprintnumbers, twocolumn, floatfix, pacs, showpacs]{revtex4}

\usepackage[dvips]{graphics}
\usepackage{amsmath,amsfonts,bm}
\usepackage{epsfig,bm}
\usepackage{graphicx,comment}

\def\cJ{{\mathcal J}}

\newcommand{\Tr}{\mbox{Tr\,}}

\newcommand{\MeV}{{\rm\ MeV}}

\newcommand{\fm}{{\rm\  fm}}

\begin{document}

\preprint{JLAB-THY-07-715}

\affiliation{Thomas Jefferson National Accelerator Facility,
             Newport News, VA 23606, USA}
\affiliation{Physics Department, Louisiana State University,
             Baton Rouge, LA 70803, USA}

\author{Hovhannes~R.~Grigoryan}
\affiliation{Thomas Jefferson National Accelerator Facility,
             Newport News, VA 23606, USA}
\affiliation{Physics Department, Louisiana State University,
             Baton Rouge, LA 70803, USA}

\newcommand\sect[1]{\noindent \textbf{\emph{#1}} --}

\title{Dimension Six Corrections to the Vector Sector of AdS/QCD Model}

\begin{abstract}
We study the effects of dimension six terms on the predictions of the holographic model for the vector meson form
factors and determine the corrections to the electric radius, the magnetic and the quadrupole moments of the
$\rho$-meson. We show that the only dimension six terms which contribute nontrivially to the vector meson form
factors are $ X^2F^2 $ and $ F^3 $. It appears that the effect from the former term is equivalent to the metric
deformation and can change only masses, decay constants and charge radii of vector mesons, leaving the magnetic
and the quadrupole moments intact. The latter term gives different contributions to the three form factors of the
vector meson and changes the values of the magnetic and the quadrupole moments. The results suggest that the
addition of the higher dimension terms improves the holographic model.
\end{abstract}


\maketitle

\sect{Introduction.} The significant progress of the holographic duals of QCD (based on \cite{Maldacena:1997re})
in determination of basic hadronic observables (see, e.g., Refs.~\cite{Erlich:2005qh}--\cite{Kwee:2007dd})
suggests for further development. In this paper, we work in the vector sector of the AdS/QCD model with the
hard-wall cutoff, proposed in the Ref.~\cite{Erlich:2005qh}. We study the effects of dimension six terms on the
vector meson form factors and extract the values of observables such as the $\rho$-meson's electric radius, the
mass, the decay constant, the magnetic and the quadrupole moments.

The leading order contribution to the vector meson form factors coming from the $ F^2 $ term has already been
studied in detail in Refs.~\cite{Grigoryan:2007my,Grigoryan:2007vg}, where it has been shown that the holographic
models in Refs.~\cite{Erlich:2005qh,Karch:2006pv} reproduce only the trivial structure of vector mesons. In
particular, instead of three independent form factors that describe vector meson, these holographic models predict
only one.

We show that the inclusion of dimension six terms changes the situation towards a more interesting scenario in
which all of the three form factors are corrected in different amounts. We also observe, that the only dimension
six terms which give nontrivial contribution to the vector meson form factors are $ X^2 F^2 $ and $ F^3 $. The
contribution from the rest of the dimension six terms can be removed by the redefinition of the coupling constant
$ g^2_5$.

We find that the addition of a term such as $ X^2 F^2 $ is equivalent to the AdS metric deformation and, according
to Ref.~\cite{Hirn:2005vk}, this, in turn, is equivalent to the inclusion of the vacuum condensates. This is in
agreement with the point made in Ref.~\cite{Erlich:2005qh} that the higher dimension (HD) operators which appear
in the operator product expansion of QCD arise in the holographic model from the higher terms in the 5D lagrangian
such as $ X^2 F^2 $. We also notice that the term $ X^2 F^2 $ doesn't alter the values of the magnetic and the
quadrupole moments, however, changes the values of the vector meson electric radius, the mass and the decay
constant.

The paper is organized as follows, in Section II, we go through the basics of the holographic model given in
Refs.~\cite{Erlich:2005qh,Erlich:2006hq}, and in particular, we discuss the leading order action, the equations of
motion for the vector bound states and the forms of dimension six terms that can enter the action. In Section III,
we demonstrate that the term like $ X^2 F^2 $ doesn't change the values of the magnetic and the quadrupole moments
and that its effect is equivalent to the AdS metric deformation. We also discuss, how this term, to a first
approximation, changes the values of the $\rho$-meson mass, the decay constant and the electric charge radius. In
Section IV, we consider the relevant part of the $ F^3 $ lagrangian and calculate the three-point function which
is then used in Section V to derive the corrections to the form factors of vector mesons. In Section VI, we
calculate the charge radius, the magnetic and the quadrupole moments of the $ \rho$-meson and compare these with
the predictions from the other models given in Refs.~\cite{Choi:1997iq}--\cite{Hedditch:2007ex}. Finally, we
summarize the paper and also show that the form factor of pion can get corrections only from the term like $ X^2
F^2 $.


\sect{Preliminaries.} We are working in the background of the sliced AdS metric of the form:
\begin{equation}\label{metric}
ds^2 = \frac{1}{z^2}\left(\eta_{\mu \nu}dx^{\mu}dx^{\nu} - dz^2\right), {\qquad}  0 < z \leq z_0 \ ,
\end{equation}
where $ \eta_{\mu \nu} = \textrm{Diag}\left(1,-1,-1,-1\right) $, $ z = z_0 $ imposes the IR hard wall cutoff,
(with $ z_0 \sim 1/\Lambda_{QCD} $) and $ z = \epsilon \rightarrow 0 $ determines the position of UV brane. From
the dictionary of the AdS/QCD model, we will correspond to the 4D vector current $J^a_{\mu}(x) =
\bar{q}(x)\gamma_{\mu}t^{a}q(x) $ a bulk gauge field $ A^a_{\mu}(x, z) $ whose boundary value is the source for $
J^a_{\mu}(x) $.  The 5D gauge action in the AdS$_5$ space is
\begin{align}\label{fullLagrangian}
S_{\rm AdS} = - \frac{1}{4g_5^2}{\rm Tr}\int d^4x~dz~\sqrt{g} \ F^{MN}F_{MN} \ ,
\end{align}
where $ F_{MN} = \partial_{M}A_{N} - \partial_{N}A_{M} - i [A_{M},A_{N}] $, $ A_{M} = t^a A^a_M $, ($ M, N = 0, 1,
2, 3, z$; \ $ \mu, \nu = 0, 1, 2, 3 $ and $t^a = \sigma^a/2 $, where $ \sigma^a $ are usual Pauli matrices with $
a = 1,2,3$). We work in the $ A_z = 0 $ gauge and require $ \partial_{\mu}A^{\mu} = 0 $.

Working in the Fourier image representation and defining $ A^a_{\mu}(q,z) = A^a_{\mu}(q)A(q,z) $, we can determine
the linearized equation of motion for $ A(q ,z) $, which is
\begin{align}\label{veceq}
\left[z^2 \partial_z^2 - z \partial_z + q^2 z^2 \right] A(q,z) = 0 \ ,
\end{align}
with boundary conditions $ A(q,0) = 1 $ and $ \partial_z A(q,z_0) = 0 $.


In general, the 5D gauge theories are not renormalizable, since the 5D gauge coupling $ g^2_5 $ has negative mass
dimension. This means that these theories can only be considered as an effective theories below some scale $
\Lambda $. In particular, for our case, the cutoff scale $ \Lambda $ should be set by $ 1/g^2_5 $.

Since, the holographic model is an effective theory with physical cutoff scale $ \Lambda \sim 1/g^2_5 $, we are
free to add HD terms into the lagrangian which respect all the required symmetries. The coefficients in front of
the dimension six operators are of the form $ c/\Lambda $, where $ c $ is some dimensionless constant and $
\Lambda = v/g^2_5 $ (it can be estimated that $ v \sim 24\pi^3 $). In general, since $ g^2_5 = 12\pi^2/N_c $,
according to Ref.~\cite{Erlich:2005qh}, we have $ c/\Lambda = 12\pi^2 c/(v N_c) \sim c/N_c $ and, therefore, for
large $ N_c $ the HD terms are $ N_c $--suppressed.

There are three groups of dimension six terms one can add into the AdS/QCD lagrangian, which may contribute to the
three-point function,
\begin{enumerate}
  \item $(\nabla_{A}F_{MN})^2 $, \ $ (\nabla_M F^{MN})^2 $ ,

  $ F^3 $, \ $ F^{MN}\nabla^2F_{MN} $ , \ $ (\nabla_KF^{MN})(\nabla_{N}F^K_{\ M})$ ,

  \item $ R F^2 $, \ $ R^{MN}F_{MK}F_{N}^{\ K} $, \ $ R^{MNKP}F_{MN}F_{KP} $ ,
  \item $ X^{\dagger}X FF $, \ $ X^{\dagger}F X F $ ,
\end{enumerate}
where $ \nabla_M $ is a covariant derivative, $ R^{MNPK} $, $ R^{MN} $ are Riemann and Ricci curvature tensors and
$ R $ is a Ricci scalar. Here, we will ignore the backreaction of the matter on the metric of the AdS space. As a
result, the contribution from the terms of the second group becomes formal, since in the AdS space these terms are
proportional to $ F^2 $ and can be absorbed into the coupling $ g^2_5 $.

Using the equation of motion
\begin{align}
\nabla_M F^{MN} = i[A_K, F^{KM}] \equiv J^M \ ,
\end{align}
it can be shown that the term $ (\nabla_M F^{MN})^2 $ doesn't contribute to the two-point and three-point
functions. Notice, that the terms $F^{MN}\nabla^2F_{MN}$ and $(\nabla_{A}F_{MN})^2 $ are equivalent, since they
differ by a full covariant derivative which vanishes after the integration because of the boundary conditions on
the fields. The terms in the third group contribute to the three-point function in such a way that the magnetic
and the quadrupole moments remain unchanged. We will show this on the example with the $ X^2 F^2 $ term.

The remaining dimension six terms which can contribute to the three-point function are given in the second line of
the first group. Using the properties of the covariant derivatives and the equation of motion, it can be shown
that
\begin{align}
F^{MN}\nabla^2F_{MN} \supset 2F^{MN}\nabla_M J_{N} \supset 2\nabla_M(F^{MN}J_{N}) \ ,
\end{align}
where we indicated only the parts which are not expressed through the terms in the second group or through the
terms which don't contribute to the three-point function. The last term enters into the action as
\begin{align}\label{HDterm1}
\Tr\int d^5 x \ & \sqrt{g} \ \nabla_M(F^{MN}J_{N})  \\ \nonumber &= - i\Tr\int d^4 x\ \left(\sqrt{g} \  F^{z
\nu}[A^{\mu}, F_{\mu \nu}]\right)_{z = 0} \ .
\end{align}
It can be shown that this term doesn't contribute to the vector meson form factors. There are different ways to
see this. One of the ways is, to notice, that the form factor is obtained as a double residue of the three-point
function (see, e.g., Ref.~\cite{Grigoryan:2007vg}). Then, working in the Fourier image representation, we have $
A(q,0) = 1 $ and, therefore, the term $[A_{\mu}, F^{\mu \nu}]_{z=0}$ can't have any poles. As can be seen from the
Eq.~(\ref{sum}), only the $ F^{z \nu} = A^{\nu}(q)\partial^z A(q,z=0) $ term in (\ref{HDterm1}), that has poles on
the UV boundary. Therefore, since, we have only one term which has poles, the double residue will vanish, leading
to zero corrections for the vector meson form factors. The similar arguments are applied for the term $ (\nabla_K
F^{MN})(\nabla_{N} F^K_{\ M})$. It appears, that only the term $ F^3 $ in this group that can give non zero
corrections to the form factors of vector mesons.

The terms of the first group $ F^{MN}\nabla^2F_{MN} $ and $ (\nabla_KF^{MN})(\nabla_{N}F^K_{\ M})$, contribute to
the two-point function only through the terms in the second group. Therefore, the effect of these terms on the two
point function is trivial and can be absorbed by the coupling $ g^2_5 $.

Notice, that the $ F^3 $ term is not coming from the expansion of DBI action. In this model, $ F^3 $ term is one
of the possible terms which should be invariant under the Lorenz and gauge transformations. We also allow the
violation of the 5D discrete charge conjugation symmetry (C) in the AdS background. As we will show, the
corrections associated with this C-violating term are $ 1/N_c^2 $ suppressed and the precise knowledge of either
the magnetic moment or the electric charge radius of the $\rho$-meson may allow to determine the holographic
bounds of strong C-violation (which is not observed in 4D).


\sect{The effects from the $ X^2 F^2 $ term.} Consider the correction to the action (\ref{fullLagrangian}), of the
form
\begin{align}
S_{X^2F^2}  &=  \kappa g^2_5 \ \Tr\int d^4 x\ dz \sqrt{g} \ X^{\dagger}X F^{MN}F_{MN} \ ,
\end{align}
where $ \kappa $ is some constant and following Ref.~\cite{Erlich:2005qh}, we have $ X^2 =
\textbf{1}_{(2\times2)}v^2(z)/4 $. In particular, $ v(z) = (m_q z + \sigma z^3) $, where $ m_q $ is the quark mass
parameter and $ \sigma $ plays the role of the chiral condensate.

We observe that the total action can be written as
\begin{align}\nonumber
S_{F^2} +  S_{X^2F^2}  &= -\frac{1}{4g^2_5}\ \Tr\int d^4 x \ dz \ \frac{p(z)}{z} \ F^{MN}F_{MN} \ ,
\end{align}
where the Lorentz indexes are now governed by the flat metric $ \eta_{MN} $, $ p(z) = 1 - \kappa g^4_5 v^2(z) $
and it is clear that, in general, the contribution from all the terms like $ X^{2n}F^2 $, ($ n $ is natural
number), will modify $ p(z) $ to a function $ P(v(z)) \equiv 1 + C_1 g^4_5 v^2(z) + \dots + C_n g^{4n}_5 v^{2n}(z)
$, where $ C_n $ are some unknown coefficients. Therefore, the inclusion of the $ X^2 F^2 $ term corresponds
effectively to the deformation of the AdS metric, that is instead of the $ 1/z^2 $ factor in the metric
(\ref{metric}), we will have $ p^2(z)/z^2 $. The similar arguments are applied also for the term $ X^{\dagger}F X
F $.

This observation allows the direct application of the result from the Ref.~\cite{Grigoryan:2007vg} to the present
case, leaving us with the following expression for the elastic form factors:
\begin{align}
\label{form1} \tilde{F}_{nn}(Q^2) = \int^{z_0}_{0}\ dz \ \frac{p(z)}{z} \, {\cal J} (Q,z) \, |\psi_n (z)|^2 \ ,
\end{align}
where $ \psi_n(z) $ are the solutions of the equations of motion,
\begin{align}\label{SL}
\partial_z\left[\frac{p(z)}{z}\partial_z \psi_n(z)\right] + \frac{p(z)}{z}M^2_n\psi_n(z) = 0 \ ,
\end{align}
with b.c. $ \psi_n(0) = \psi'_n(z_0) = 0 $ and $ q^2 = M^2_n $. The function $ \cJ(Q,z) $ is a solution of the
same equation of motion but with $ q^2 = - Q^2 $ instead of $ M_n^2 $ and b.c. $ \cJ(Q,0) = 1 $, $ \partial_z
\cJ(Q,z_0) = 0 $. The eigenfunctions of Eq.~(\ref{SL}) are normalized as
\begin{align}
\int^{z_0}_{0}\ dz \ \frac{p(z)}{z} |\psi_n (z)|^2 = 1 \ .
\end{align}
Therefore, $ \tilde{F}_{nn}(0) = 1 $ and, since, the electric $G_C$, magnetic $G_M$ and quadrupole $G_Q$ form
factors are:
\begin{align}\nonumber
&G_Q ^{(n)}(Q^2) = - \tilde{F}_{nn}(Q^2) \ , \ \ G_M^{(n)} (Q^2) = 2
\tilde{F}_{nn}(Q^2)\ , \\[7pt] \label{GCff}  &G_C^{(n)} (Q^2) = \left  (1 -
\frac{Q^2}{6M^2_n} \right )\tilde{F}_{nn}(Q^2) \ ,
\end{align}
one can check that at $ Q^2 = 0 $, these form factors reproduce the same values for electric charge, magnetic and
quadrupole moments, as in the case with $ \kappa = 0 $, that is in the absence of the $ X^2 F^2 $ term. This term,
however, can change masses and decay constants of vector mesons. Besides, it also changes the electric radius of
the $\rho$-meson.


Notice, that the eigenvalues of the Eq.~(\ref{SL}) may be expressed through the eigenfunctions in the following
way:
\begin{align}
M^2_{n} = \int^{z_0}_{0}\ dz \ \frac{p(z)}{z} |\partial_z \psi_n (z)|^2 \ .
\end{align}
Up to a first order approximation, using the same eigenfunctions as in case with $ \kappa = 0 $, that is
\begin{equation}\label{holowf}
\psi^{(0)}_n(z) = \frac{\sqrt{2}}{z_0J_1(\gamma_{0,n})} \ z J_1(M^{(0)}_{n}z) \ ,
\end{equation}
with $ M^{(0)}_{n} = \gamma_{0,n}/z_0 $ (where $ J_0(\gamma_{0,n}) = 0 $) but with metric perturbation $ p(z) $,
we will have for the $\rho$-meson mass $ M_{\rho} \equiv M_1 $ the following result:
\begin{align}\label{massf1}
M_{\rho} \simeq M^{(0)}_{\rho}\left(1 - 0.02\kappa g^4_5\right) \ ,
\end{align}
where $ M^{(0)}_{\rho} $ is the mass of the $\rho$-meson in case $ \kappa = 0 $, and we used the values of
parameters: $ m_q = 2.3 \MeV $, $ \sigma = (327 \MeV)^3 $, $ z_0 = 1/(323 \MeV) $, taken from the Model A of
Ref.~\cite{Erlich:2005qh}.

The decay constant of the $ \rho$-meson, $ f_{\rho} $, in terms of the eigenfunctions of the 5D equation of motion
has the form
\begin{align}
f_{\rho} = \frac{1}{g_5}\left(\frac{p(z)}{z}\partial_z\psi_{\rho}(z)\right)_{z \rightarrow 0} \ ,
\end{align}
as was discussed, for example, in the Ref.~\cite{Grigoryan:2007my}. The solution for $ \psi_{\rho}(z) \equiv
\psi_1(z) $ near the $ z = 0 $ is of the same form as in case $ \kappa = 0 $ thus,
\begin{align}
f_{\rho} = \frac{\sqrt{2}M_{\rho}}{g_5 z_0 J_1(\gamma_{0,1})} \ .
\end{align}
Therefore, to lowest order in $ \kappa $, we will have:
\begin{align}\label{decay1}
f_{\rho} \simeq f^{(0)}_{\rho}\left(1 - 0.02\kappa g^4_5\right) \ ,
\end{align}
where $ f^{(0)}_{\rho} $ is the decay constant in case when $ \kappa = 0 $.

We can also express the electric charge radius of the $\rho$-meson, $ \langle \tilde{r^2_{\rho}} \rangle_C  $,
defined as
\begin{align}\label{raddef}
\langle \tilde{r^2_{\rho}} \rangle_C &\equiv -6 \left(\frac{dG^{(1)}_C(Q^2)}{dQ^2}\right)_{Q^2 = 0} \ ,
\end{align}
in terms of the parameter $ \kappa $. In this case, using the Eqs.~(\ref{form1}), (\ref{GCff}) and (\ref{raddef}),
to lowest order in the coefficient $ \kappa $, the electric charge radius is:
\begin{align}\label{rad1}
\langle \tilde{r^2_{\rho}} \rangle_C \simeq (0.53 - 0.16\kappa g^4_5) \fm^2 \ ,
\end{align}
where $ 0.53 \fm^2 $ is the result for the electric radius obtained in Ref.~\cite{Grigoryan:2007vg} (again, we
used parameters taken from the Model A of Ref.~\cite{Erlich:2005qh}).

The similar analysis can be applied for the case of Model B in Ref.~\cite{Erlich:2005qh}, for which we have:
\begin{align}\label{values}
M_{\rho} &\simeq M^{(0)}_{\rho}\left(1 - 0.01\kappa g^4_5\right) \ , \\ \nonumber
f_{\rho} &\simeq f^{(0)}_{\rho}\left(1 - 0.01\kappa g^4_5\right) \ , \\ \nonumber
\langle \tilde{r^2_{\rho}} \rangle_C  &\simeq (0.46 - 0.07\kappa g^4_5) \fm^2 \ .
\end{align}
Notice, that the coefficients in front of $ \kappa $, in case of Model B are almost twice as smaller than in the
Model A. Also, it is straightforward to see that the contribution from the term $ X^{\dagger}F X F $ can be
absorbed by $ \kappa $.

Now, since $ g^2_5 = 12\pi^2/N_c $, it follows that the corrections to the observables ($ \sim \kappa g^4_5 $) are
$ 1/N^2_c $ suppressed. The natural constraint on the coefficient $ \kappa $ should come from the requirement that
the corrections to the observables are small. This means that, if $ N_c = 3 $, then for the first two observables
in (\ref{values}), we should have $ |\kappa| \ll 0.06 $ and for the third one we expect to have $ |\kappa| \ll
0.004 $. Therefore, we conclude, that it is natural for the coefficient $ \kappa $ to satisfy the condition $
|\kappa| \ll 10^{-3} $.


\sect{Corrections from the $ F^3 $ term.} The action relevant for finding the corrections to the 3-point function
is
\begin{align}
S_{F^3} &= \alpha g^2_5 \Tr\int d^4x~dz~\sqrt{g}\left(F_{MN}F^{NK}F_K^{\ M} \right) \\ \nonumber &\supset
\frac{i\alpha g^2_5 \epsilon^{abc}}{4}\int d^4x~dz~z~\biggl[3(\partial_{\mu} A^{a}_{\nu})(\partial_{z}
A^{b,\nu})(\partial_{z} A^{c,\mu})\\ \nonumber & {\qquad} {\qquad} \ \ \ {\qquad} {\qquad}{\qquad} +
2(\partial^{\mu} A^{a,\nu})F_{\nu}^{b,\alpha}F_{\alpha\mu}^c\biggr] \ ,
\end{align}
where $ \alpha $ is a new dimensionless (C-violating) parameter of the theory and the Lorentz indexes are governed
by the Minkowski flat metric $ \eta_{\mu \nu} $. Therefore, using the prescription of the holographic model, for
the 3-point function we will have:
\begin{align}
&T^{abc}_{\mu\alpha\beta}(p_1,p_2,q) \equiv \langle J^b_{\alpha}(p_1) J^a_{\mu}(q)J^c_{\beta}(-p_2)\rangle \\[7pt]
\nonumber  &= \epsilon^{abc}T_{\mu\alpha\beta}(p_1,p_2,q)i(2\pi)^4\delta^{(4)}(q - p_2 + p_1) \ ,
\end{align}
where
\begin{align}
\nonumber T_{\mu\alpha\beta}(p_1,p_2,q) &= \frac{3\alpha g^2_5}{4}\biggl\{ \left[q^2K_2 -
K_{11}\right]\eta_{\alpha\beta}(p_1 + p_2)_{\mu} \\[5pt] \nonumber &+ \left[2M^2K_2 -
K_{12}\right](\eta_{\mu\alpha}q_{\beta} - \eta_{\mu\beta}q_{\alpha}) \\[6pt] &- 2K_2q_{\alpha}q_{\beta}(p_1 +
p_2)_{\mu}\biggr\} \ ,
\end{align}
and
\begin{align}
\nonumber K_{11}(p_1,p_2,q ) &= \int^{z_0}_{0}dz \ z \partial_{z} A(q,z)A(p_1,z)\partial_zA(p_2,z)  \ , \\
\nonumber K_{12}(p_1,p_2,q ) &= \int^{z_0}_{0}dz \ z \partial_{z}\left[A(q,z)A(p_1,z)\right]\partial_zA(p_2,z)  \ , \\
K_2(p_1,p_2,q ) &= \int^{z_0}_{0}dz \ z A(q,z)A(p_1,z)A(p_2,z) \ ,
\end{align}
where we used that the functions $ K(p_1,p_2,q) $ are symmetric under the exchange of $ p_1 \leftrightarrow p_2 $
(to understand this, see Eq.~(\ref{sum})), but not $ p_{1,2} \leftrightarrow q $, ($ q = p_2 - p_1 $) and
anticipating the on-shell limit, we applied conditions: $ p^2_1 = p^2_2 = M^2 $, $ (p_1p_2) = M^2 - q^2/2 $ and $
(p_2q) = -(p_1q) = q^2/2 $, for the diagonal transitions (one can easily generalize this to non diagonal
transition). Since we are dealing with the transverse components of the gauge field, to simplify the tensor
structure, we applied, as in \cite{Grigoryan:2007my}, the transverse projectors $ \Pi^{\alpha\alpha'}(p_1) \equiv
(\eta^{\alpha\alpha'} - p_{1}^{\alpha}p_{1}^{\alpha'}/p^2_1) $, etc, (that allows us to add or eliminate terms
proportional to $ p_{1\alpha} $ or $ p_{2\beta} $). The solution of the (\ref{veceq}) for timelike momentum can be
written as an infinite sum:
\begin{equation}\label{sum}
A(p,z) = -g_5 \sum_{m = 1}^{\infty}\frac{f_{m}\psi_m(z)}{p^2 - M^2_{m} } \ ,
\end{equation}
where $ \psi_m(z) $ are the solutions of the (\ref{veceq})  with b.c. $ \psi_m(0) = \psi'_m(z_0) = 0 $ and $ q^2 =
M^2_m $. Then, for a spacelike  momentum transfer, $ q^2 = - Q^2 $, it follows that:
\begin{align}\nonumber
T_{\mu\alpha\beta}(p_1,p_2,q) = \frac{3\alpha g^4_5}{4}\sum^{\infty}_{n,k = 1} \frac{f_m f_n
R^{nk}_{\mu\alpha\beta}(Q^2)}{(p_1^2 - M_n^2)(p^2_2 - M_k^2)} \ ,
\end{align}
and for the diagonal $ n \leftrightarrow n $ transition:
\begin{align}
\nonumber R^{(n)}_{\mu\alpha\beta}(Q^2) &\equiv \lim_{p^2_1 \rightarrow M^2_n}\lim_{p^2_2 \rightarrow M^2_n}(p^2_1
- M^2_n)(p^2_2 - M^2_n)
T_{\mu\alpha\beta} \\[7pt] \nonumber &=
\frac{3\alpha g^4_5}{4}\biggl\{-\left[Q^2W^{nn}_2 + W^{nn}_{11}\right]\eta_{\alpha\beta}(p_1 + p_2)_{\mu}  \\[7pt]
\nonumber  &+ \left[2M^2_nW^{nn}_2 - W^{nn}_{12}\right](\eta_{\mu\alpha}q_{\beta} - \eta_{\mu\beta}q_{\alpha}) \\[7pt]
&- 2W^{nn}_2q_{\alpha}q_{\beta}(p_1 + p_2)_{\mu}\biggr\} \ ,
\end{align}
where we defined new functions as
\begin{align} \label{w11term}
W^{nn}_{11}(Q^2) &= \int^{z_0}_{0}dz \ z \partial_{z}\cJ(Q,z)\psi_n(z)\partial_z\psi_n(z)  \ , \\
W^{nn}_{12}(Q^2) &= \int^{z_0}_{0}dz \ z \partial_{z}\left[\cJ(Q,z)\psi_n(z)\right]\partial_z\psi_n(z)  \ , \\
W^{nn}_2(Q^2) &= \int^{z_0}_{0}dz \ z \cJ(Q,z)\psi_n(z)\psi_n(z) \ .
\end{align}
with
\begin{align}
\cJ(Q,z) &= {Qz} \left[ K_1(Qz) + I_1(Qz) \frac{K_0(Qz_0)}{I_0(Qz_0)} \right] \ ,
\end{align}
where $ \cJ(Q,z) = A(Q,z) $ is the solution of Eq.~(\ref{veceq}).


\sect{Form Factors.} Adding the corrections to the form factor coming from the $ F^3 $ term to the leading order
result from the $ F^2 $ term obtained in Ref.~\cite{Grigoryan:2007vg} gives for the electric $\tilde{G}_C$,
magnetic $\tilde{G}_M$ and quadrupole $\tilde{G}_Q$ form factors the following result
\begin{align}
\nonumber \tilde{G}^{(n)}_C(Q^2) &= \left[1-\frac{Q^2}{6M^2_n}\right]F_{nn} - \frac{3\alpha g^4_5 Q^2}{4}\left[1 +
\frac{Q^2}{12M^2_n}\right]W^{nn}_2\\ \nonumber &- \frac{3\alpha g^4_5}{4}\left[1  +
\frac{Q^2}{6M^2_n}\right]W^{nn}_{11}
+ \frac{\alpha g^4_5 Q^2}{8M^2_n}W^{nn}_{12} \ ,\\[7pt]
\nonumber
\tilde{G}^{(n)}_M(Q^2) &= 2F_{nn}(Q^2) + \frac{3\alpha g^4_5}{4}\left[2M^2_nW^{nn}_2 - W^{nn}_{12}\right] \ , \\
\nonumber \tilde{G}^{(n)}_Q(Q^2) &= - F_{nn}(Q^2) - \frac{3\alpha g^4_5 Q^2}{8}W^{nn}_2 \\  &- \frac{3\alpha
g^4_5}{4}\left[W^{nn}_{11} - W^{nn}_{12}\right] \ .
\end{align}
where
\begin{align}
F_{nn}(Q^2) = \int^{z_0}_{0} \frac{dz }{z} \,  {\cal J} (Q,z) \, |\psi_n (z) |^2 \ ,
\end{align}
see Ref.~\cite{Grigoryan:2007vg} for more details.
In the AdS/QCD model, with $ \alpha = 0 $ as was shown in \cite{Grigoryan:2007vg}, these three form factors of
vector meson are expressed through the single function $ F_{nn}(Q^2) $. Besides, for $Q^2=0$, the AdS/QCD model
reproduce the unit electric charge $ e $ of the meson, ``predict'' \mbox{$\mu \equiv G_M(0) = 2$} for the magnetic
moment and \mbox{$D\equiv G_Q(0)/M^2 = -1/M^2 $} for the quadrupole moment, which are just the canonical values
for a vector particle \cite{Brodsky:1992px}. However, for non zero value of $ \alpha $ the situation changes
towards a more realistic scenario.


\sect{Results.} One can verify that at $ Q^2 = 0 $, we have $ W^{nn}_{11}(0) = 0 $, because $
\partial_z \cJ(0,z) = 0 $, since
\begin{equation}
\label{JD0} \partial_z  \, {\cal J} (Q,z) = - z{Q^2}\biggl[ K_0(Qz) -  I_0(Qz) \,  \frac{K_0(Qz_0)}{ I_0(Qz_0)}
\biggr ] \ .
\end{equation}
Besides
\begin{align}
W^{11}_{12}(0) &= \int^{z_0}_{0}dz \ z (\partial_{z}\psi_1(z))^2 \\ \nonumber
&= \frac{2M^2z^2_0}{J^2_1(\gamma_{0,1})}\int^{1}_{0}d\zeta \ \zeta^3 J^2_0(\gamma_{0,1} \zeta)  \ , \\[7pt]
W^{11}_2(0) &= \int^{z_0}_{0}dz \ z \psi^2_1(z) \\ \nonumber &= \frac{2z^2_0}{
J^2_1(\gamma_{0,1})}\int^{1}_{0}d\zeta \ \zeta^3 J^2_1(\gamma_{0,1}\zeta) \ ,
\end{align}
where $ J_0(\gamma_{0,1}) = 0 $, $ M = \gamma_{0,1}/z_0 $ is the mass of the $\rho$-meson and we took into account
that
\begin{align}\label{wf}
\psi_1(z) = \frac{\sqrt{2}}{z_0J_1(\gamma_{0,1})} \ zJ_1(M z) \ .
\end{align}
After partial integrations and using the properties of Bessel functions we will have
\begin{align}
W^{11}_{12}(0) = M^2W^{11}_2(0) - 2 \ .
\end{align}
Now, defining $ w \equiv W^{11}_{12}(0) \simeq 1.261 $, we find ($ e = 1 $),
\begin{align}
\mu &\equiv \tilde{G}^{(1)}_M(0) =  2 + \frac{3\alpha g^4_5}{4}\left(w + 4\right)  \ , \\
\nonumber D M^2 &\equiv \tilde{G}^{(1)}_Q(0) = - 1 + \frac{3\alpha g^4_5 w}{4}  \ .
\end{align}
The electric radius of the $\rho$-meson is
\begin{align}
&\langle \tilde{r^2_{\rho}} \rangle_C \equiv -6\left(\frac{d\tilde{G}^{(1)}_C(Q^2)}{dQ^2}\right)_{Q^2 = 0} = \langle r^2_{\rho} \rangle_C \\[5pt] \nonumber
&+ \alpha g^4_5\biggl[\frac{3}{4M^2}(5w+12) + \frac{9}{2}\left(\frac{dW^{11}_{11}(Q^2)}{dQ^2}\right)_{Q^2 =
0}\biggr] \ ,
\end{align}
where the first term is $ \langle r^2_{\rho} \rangle_C = 0.53 \fm^2 $, found in Ref.~\cite{Grigoryan:2007vg}, and
the second term in the square brackets is the correction to the $\rho$-meson's radius. Using the
Eqs.~(\ref{w11term}), (\ref{JD0}) and (\ref{wf}) one can find that
\begin{align}
&\frac{9}{2}\left(\frac{dW^{11}_{11}(Q^2)}{dQ^2}\right)_{Q^2 = 0} = \\[5pt] \nonumber &=
\frac{9\gamma_{0,1}z_0^2}{J^2_1(\gamma_{0,1})}\int^1_0 d\zeta \ \zeta^4 \ln \zeta \ J_0(\gamma_{0,1}\zeta)
J_1(\gamma_{0,1}\zeta) \ ,
\end{align}
which is  $ \simeq -0.255 \fm^2 $. Therefore,
\begin{align}
&\sigma \equiv \left(\langle \tilde{r^2_{\rho}} \rangle_C - \langle r^2_{\rho} \rangle_C\right)/\fm^2 \simeq 0.647
\alpha g^4_5 \simeq 252 \alpha  \ .
\end{align}
Now, in terms of $ \sigma $, the magnetic and quadrupole moments of the $\rho$-meson are: $ \mu \simeq 2 +
6.1\sigma $ and $ DM^2 = 1.46\sigma - 1 $. The table of possible values for electric radius, magnetic and
quadrupole moments in terms of reasonable range of values for $ \sigma $ is given below:
\begin{center}
\begin{tabular}[t]{|l|c|c|c|c|c|c|c|c|}
\multicolumn{9}{c}{TABLE I: The observables for different values of $ \sigma $.}\\ \hline $ \ \ \ \sigma$ &-0.15 &-0.1 &-0.05 &-0.01 &0.01 &0.05 & 0.1 & 0.15 \\
\hline \hline
$ \ \ \ r^2 $ &0.38&0.43 & 0.48 & 0.52 & 0.54 & 0.58 & 0.63 & 0.68 \\
$ \ \ \ \mu$ & 1.09 & 1.39 & 1.7 & 1.94 & 2.06 & 2.31 & 2.61 & 2.92 \\
$-DM^2$ & 1.22 & 1.15 & 1.07 & 1.01 & 0.99 & 0.93 & 0.85 & 0.78 \\ \hline
\end{tabular}
\end{center}
where $r^2 \equiv \langle \tilde{r^2_{\rho}}\rangle_C/\fm^2 $. These results depend explicitly on $ \alpha $ (or $
\sigma $) and implicitly on $ z_0 $ which is fixed by the mass of the $\rho$-meson. Notice, that $ g^4_5|\alpha| <
0.23 $, therefore, we are not outside of the perturbative domain and our calculations are consistent. For
comparison with other models, see table below
\begin{center}
\begin{tabular}[t]{|l|c|c|c|c|c|c|}
\multicolumn{7}{c}{TABLE II: The observables in different models.}\\ \hline \ \ Models \ \ & \ \cite{Choi:1997iq}
\ & \ \cite{Burden:1995ve} \ & \ \cite{deMelo:1997hh} \ & \ \cite{Frankfurt:1993ut} \ & \ \cite{Maris:1999bh} \ &
\cite{Hedditch:2007ex} \\ \hline \hline
$ \ \ \ r^2 $ & 0.27 & 0.37 & 0.37 & 0.39 & 0.54 & 0.55 \\
$ \ \ \ \mu$ & 1.92 & 2.69 & 2.14 & 2.48 & 2.01 & 2.25 \\
     $-DM^2$ & 0.43 & 0.84 & 0.79 & 0.89 & 0.41 & 0.11 \\ \hline
\end{tabular}
\end{center}
It is interesting, that the only HD term in the 5D effective theory that can alter the canonical values of the
magnetic and the quadrupole moments is the C-violating term $ F^3 $. Therefore, the more precise knowledge of
either one of these observables ($\mu$, $D$ or $ r^2 $) can put more stringent constraints on the C-violating
coefficient $ \alpha $. Here, we showed that the corrections are proportional to $ \alpha g^4_5 $ and thus, are $
1/N^2_c $ suppressed as expected. Finally, our estimates suggest that $ |\alpha| < 10^{-4} $.


\sect{Summary.} In this paper, as one of the possible ways to test and improve the AdS/QCD model proposed in the
Ref.~\cite{Erlich:2005qh}, we considered the addition of dimension six terms into the vector sector of the AdS/QCD
lagrangian and study their effect on the vector meson form factors.

We discussed that ignoring the backreaction of the matter to the metric, the effect from the terms of the second
group involving the AdS curvature tensors and Ricci scalar, is equivalent to the redefinition of the coupling $
g^2_5 $. We showed that the term, like $ X^2 F^2 $, doesn't change the electric charge, the magnetic and the
quadrupole moments, but affects the charge radius, the masses and the decay constants of the vector mesons. The
effect of this term is equivalent to the AdS metric deformation and, in agreement with \cite{Erlich:2005qh} and
\cite{Hirn:2005vk}, it is also equivalent to the addition of the vacuum condensates. However, one should keep in
mind that the metric deformations are also coming from the matter fields, which we ignore compared to the explicit
or effective metric deformations from the $ X^2F^2$ term.

By calculating the form factors, we found a relation between electric charge radius, mass and decay constant of
the $\rho$-meson on the coefficient $ \kappa $ (to lowest order) with which the term $ X^2 F^2 $ enters the
action. Also, we expressed electric radius, magnetic and quadrupole moments of the $\rho$-meson in terms of the
dimensionless parameter $ \alpha $, with which the term $ F^3 $ enters the action. These results can be
straightforwardly generalized to the case of the soft wall model \cite{Karch:2006pv,Grigoryan:2007my}.

It is also interesting to study the contribution of the dimension six terms to the form factor of pion. As it was
discussed in Ref.~\cite{Grigoryan:2007wn}, in the full AdS/QCD model the pion form factor is derived from the
variation of the action with respect to the two longitudinal axial-vector fields and one transverse vector field.
As a result, only the term like $ X^2 F^2 $ can contribute to the form factor of pion. To demonstrate this, first,
consider the term $ F^2_AF_V $, where $ F_A $ is related to the axial-vector field. This term may contribute to
the three-point function in such a way that only the linear pieces of the field strength tensors can enter.
However, since these linear pieces vanish for the longitudinal axial-vector field, there can't be any contribution
from the term like $ F^3 $ to the form factor of pion (this question was also discussed in
Ref.~\cite{Kwee:2007dd}).

The other relevant dimension six terms $ (\nabla_{A}F_{MN})^2 $ and $ (\nabla_KF^{MN})(\nabla_{N}F^K_{\ M}) $ also
can't contribute to the form factor of pion. We demonstrate this on the example with the term $
(\nabla_{A}F_{MN})^2 $ which, as shown above contributes to the action in the form given in Eq.~(\ref{HDterm1}).
However, this term contains two field strength tensors, and at least one should vanish for the longitudinal
components. Similar arguments can  be also applied for the second term.

Finally, we think that the results obtained here are in the range of the values from the other models. This is
encouraging and suggests that the further addition of the HD terms can improve the holographic dual model of QCD.


\sect{Acknowledgments.} I would like to thank A.V. Radyushkin, J. Goity, J. Erlich and C. Carone for helpful
discussions, A.W. Thomas for valuable comments and support at Jefferson Laboratory and J.P. Draayer for support at
Louisiana State University.

Notice: Authored by Jefferson Science Associates, LLC under U.S. DOE Contract No. DE-AC05-06OR23177. The U.S.
Government retains a non-exclusive, paid-up, irrevocable, world-wide license to publish or reproduce this
manuscript for U.S. Government purposes.


\end{document}